\documentclass[useAMS,usenatbib,usegraphicx]{mn2e}
\usepackage{longtable}
\usepackage{lscape}
\usepackage{times}
\usepackage{url}

\title[Fe~I Oscillator Strengths for the Gaia-ESO Survey]{Fe~I Oscillator Strengths for the Gaia-ESO Survey}
\author[M. P. Ruffoni, E. A. Den Hartog, J. E. Lawler, N. R. Brewer, K. Lind, G. Nave and J. C. Pickering]
{\parbox{\textwidth}{M. P. Ruffoni$^{1}$\thanks{E-mail:m.ruffoni@imperial.ac.uk},
E. A. Den Hartog$^{2}$,
J. E. Lawler$^{2}$,
N. R. Brewer$^{2}$,
K. Lind$^{3}$,
G. Nave$^{4}$ and 
J. C. Pickering$^{1}$}\vspace{0.2cm}\\
$^{1}$Blackett Laboratory, Imperial College London, London SW7 2BW, UK\\
$^{2}$Department of Physics, University of Wisconsin, Madison, WI 53706, USA\\
$^{3}$Institute of Astronomy, University of Cambridge, Cambridge CB3 0HA, UK\\
$^{4}$National Institute of Standards and Technology, Gaithersburg, Maryland 20899-8422, USA}
\begin{document}

\date{}

\pagerange{\pageref{firstpage}--\pageref{lastpage}} \pubyear{2013}

\maketitle

\label{firstpage}

\begin{abstract}
The Gaia-ESO Public Spectroscopic Survey (GES) is conducting a large-scale study of multi-element chemical abundances of some $100~000$ stars in the Milky Way with the ultimate aim of quantifying the formation history and evolution of young, mature and ancient Galactic populations. However, in preparing for the analysis of GES spectra, it has been noted that atomic oscillator strengths of important Fe~I lines required to correctly model stellar line intensities are missing from the atomic database. Here, we present new experimental oscillator strengths derived from branching fractions and level lifetimes, for 142 transitions of Fe~I between 3526~\AA~and 10864~\AA, of which at least 38 are urgently needed by GES. We also assess the impact of these new data on solar spectral synthesis and demonstrate that for 36 lines that appear unblended in the Sun, Fe abundance measurements yield a small line-by-line scatter (0.08 dex) with a mean abundance of 7.44 dex in good agreement with recent publications.
\end{abstract}


\section{Introduction}

The Gaia-ESO Public Spectroscopic Survey (GES) is currently taking place at the European Southern Observatory (ESO), employing the Fibre Large Array Multi Element Spectrograph (FLAMES) instrument at the Very Large Telescope (VLT) facility. Its aim is to obtain high quality spectroscopy of some 100~000 stars from all major components of the Milky Way to quantify the ``kinematic multi-chemical element abundance distribution functions of the Milky Way Bulge, the thick Disc, the thin Disc, and the Halo stellar components, as well as a very significant sample of 100 open clusters" \citep{ref:gilmore12}. Over the course of the survey, chemical abundances will be measured for alpha and iron-peak elements in all stars with visual magnitude less than nineteen. These data will probe stellar nucleosynthesis by examining nuclear statistical equilibrium and the alpha-chain. Ultimately, the abundances and radial velocities will be combined with high-precision position and proper motion measurements from the European Space Agency's \emph{Gaia} mission, to ``quantify the formation history and evolution of young, mature and ancient Galactic populations" \citep{ref:perryman01}. \cite{ref:gilmore12} also state that ``Considerable effort will be invested in abundance calibration and ESO archive re-analysis to ensure maximum future utility."

To achieve these high-level aims, it is vital that fundamental atomic data be available for lines in the GES spectral range: $4800$ \AA~to $6800$ \AA~ for measurements with the high-resolution FLAMES Ultraviolet and Visual Echelle Spectrograph (UVES) and $8500$ \AA~to $9000$ \AA~for measurements with the mid-resolution FLAMES \emph{Giraffe} spectrograph. The availability of absorption oscillator strengths, f (usually used as the $\log(gf)$, where $g$ is the statistical weight of the lower level), is particularly important for the correct modelling and analysis of stellar line intensities; especially so for abundant elements such as iron, which is also used to infer fundamental stellar parameters. 

However, in preparing a list of iron lines to be targeted during the analysis of GES spectra, the GES line list team noted that of $449$ well-resolved lines of neutral iron (Fe~I) expected to be visible with sufficient signal-to-noise ratio, only 167 have published log(gf) values measured in the laboratory with uncertainties below 25 \%. Experimental log(gf) values with large uncertainties (greater than 50 \% in many cases) were available for an additional 162 lines. For the final $120$ lines, no experimental $\log(gf)$s were available at all. A similar observation was made by \cite{ref:bigot06} for lines of interest to the \emph{Gaia} mission.

As a result of this inadequacy in the atomic database, and similar inadequacies observed by other astronomers (see \cite{ref:ruffoni13b} and \cite{ref:pickering11}, for example), we have undertaken a new study of the Fe~I spectrum with the aim of providing accurate $\log(gf)$ values for lines of astrophysical significance. In Section \ref{section:results} of this paper, we report accurate $\log(gf)$s for 142 Fe~I lines, 64 of which have been measured experimentally for the first time. The $\log(gf)$ values of at least 38 of these lines are urgently needed for the GES survey. 

\section{Experimental Procedure}\label{section:expt}

Typically, $\log(gf)$s are obtained in the laboratory from measurements of atomic transition probabilities, A, \citep{ref:spectrophysics}.
\begin{equation}
\log(gf) = \log\Bigl[A_{ul}g_u \lambda^2\times 1.499\times 10^{-14} \Bigr]~,
\end{equation}
where the subscript $u$ denotes a target upper energy level, and $ul$, a transition from this level to a lower level, $l$, that results in emission of photons of wavelength $\lambda$ (nm). $g_u$ is the statistical weight of the upper level. The $A_{ul}$ values are found by combining experimental branching fractions, $\mbox{BF}_{ul}$, with radiative lifetimes, $\tau_u$ \citep{ref:huber86}.
\begin{equation}
\label{eqn:trprob}
A_{ul} = \frac{\mbox{BF}_{ul}}{\tau_u}~;~~ \tau_u = \frac{1}{\sum_l A_{ul}}.
\end{equation}
The BF$_{ul}$ for a given transition is the ratio of its $A_{ul}$ to the sum of all $A_{ul}$ associated with $u$. This is equivalent to the ratio of observed relative line intensities in photons/s for these transitions.
\begin{equation}
\label{eqn:bf}
\mbox{BF}_{ul} = \frac{A_{ul}}{\sum_lA_{ul}} = \frac{I_{ul}}{\sum_lI_{ul}}~.
\end{equation}
This approach does not depend on any form of equilibrium in the population distribution over different levels, but it is essential that all significant transitions from $u$ be included in the sum over $l$.

The BFs measured for this work were extracted from Fe~I spectra acquired by Fourier transform (FT) spectroscopy, as described in Section \ref{section:bf}. The radiative lifetimes required to solve Equation \ref{eqn:trprob} were obtained through laser induced fluorescence (LIF), and are discussed in Section \ref{section:lif}.

\subsection{Branching Fraction Measurements}
\label{section:bf}
The BFs reported here were obtained from Fe~I emission line spectra measured in two overlapping spectral ranges between $8200$ cm$^{-1}$ and $35500$ cm$^{-1}$ (between $1220$ nm and $282$ nm), labelled A and B in Table \ref{table:spectra}. 

Spectrum A was measured between $8200$ cm$^{-1}$ and $25500$ cm$^{-1}$ (3920.5 \AA~and 12191.8 \AA) on the 2 m Fourier transform (FT) spectrometer at the National Institute of Standards and Technology (NIST) \citep{ref:xgremlin}. The Fe I emission was generated from an iron cathode mounted in a water cooled hollow cathode lamp (HCL) running at a current of 2.0 A in a Ne atmosphere of 370 Pa pressure. The response of the spectrometer as a function of wavenumber was obtained by measuring the spectrum of a calibrated tungsten (W) halogen lamp with spectral radiance known to $\pm$1.1 \% between $250$ nm and $2400$ nm. W lamp spectra were acquired both before and after measurements of the Fe/Ne HCL spectrum to verify that the spectrometer response remained stable.

\begin{figure*}
\centering
\includegraphics[scale=0.5]{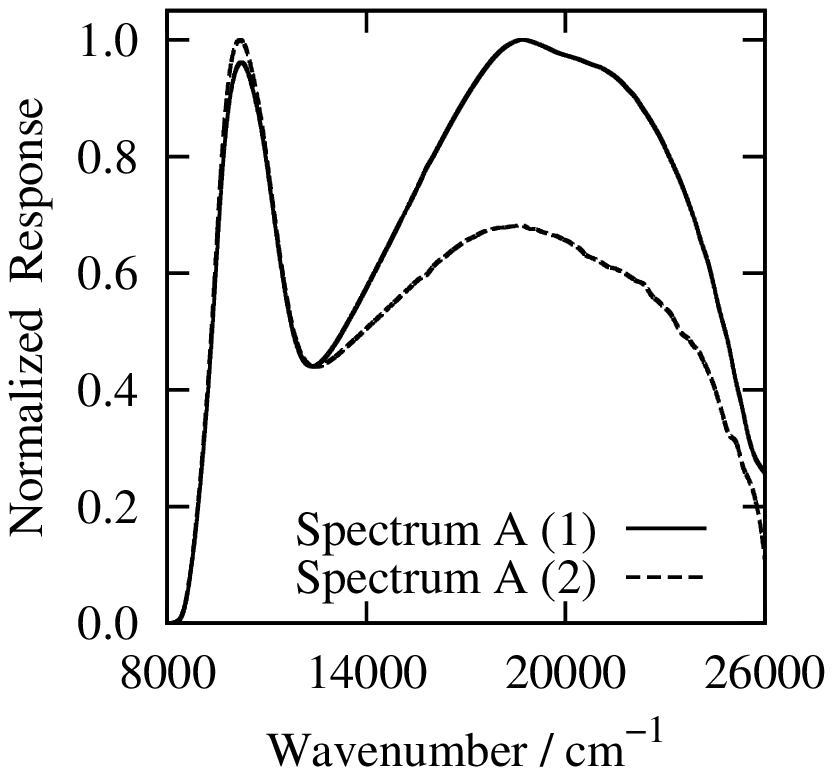}\includegraphics[scale=0.5]{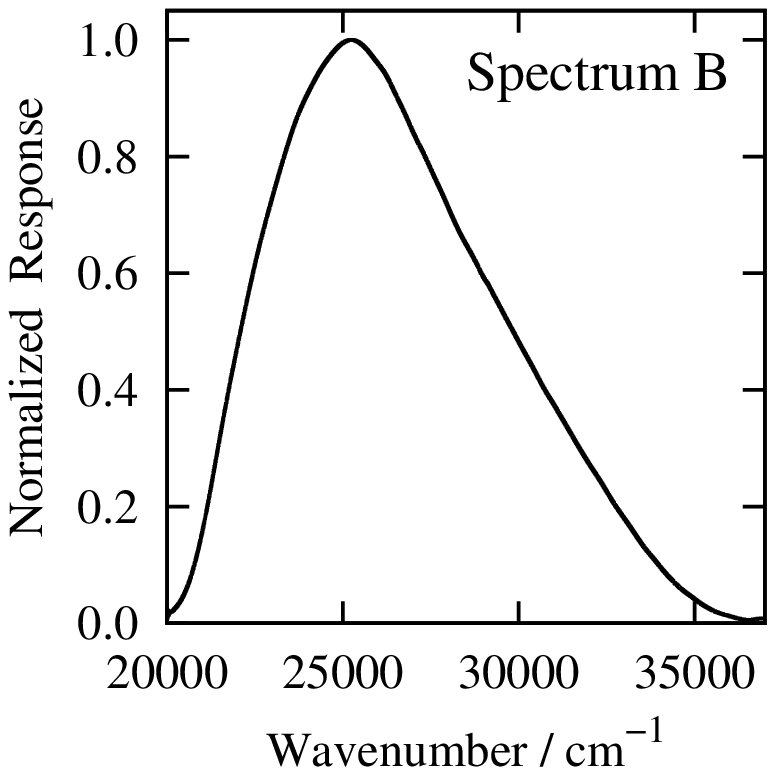}
\caption{Instrument response functions used to intensity calibrate the two FT spectra listed in Table \ref{table:spectra}.}
\label{fig:response}
\end{figure*}

220 individual Fe/Ne HCL spectra were acquired over two days and coadded to improve the signal-to-noise ratio of weak lines. However, due to different detector configurations being used on each day, the spectrometer response function varied significantly between the two. As a result, the files Fe080311.001 to .003 (containing 110 spectra, acquired with a Si photodiode on each outputs of the FT spectrometer) were coadded and intensity calibrated using the spectral response function labelled ``Spectrum A (1)" in Figure \ref{fig:response}, while Fe080411\_B.001 to .003 (containing the remaining 110 spectra, acquired with a single Si photodiode mounted on the unbalanced output of the FT spectrometer) were coadded and calibrated using the response function labelled ``Spectrum A (2)". These response functions were obtained with the aid of the \verb|FAST| package \citep{ref:ruffoni13}. The two intensity calibrated line spectra were then themselves coadded to produce the final spectrum.

Spectrum B was measured between $20000$ cm$^{-1}$ and $35500$ cm$^{-1}$ (2816.1 \AA~and 4998.6 \AA) on the Imperial College VUV spectrometer \citep{ref:thorne96}, which is based on the laboratory prototype designed by \cite{ref:thorne87}. The Fe emission was generated from an iron cathode mounted in a new HCL designed and manufactured at Imperial College London (IC). The lamp was operated at 700 mA in a Ne atmosphere of 170 Pa pressure to provide reasonable signal-to-noise ratio in the weaker lines while avoiding self-absorption effects in the stronger lines. 

The spectrometer response function for spectrum B is also shown in Figure \ref{fig:response}, and was again obtained from a calibrated W lamp, measured before and after each Fe/Ne HCL measurement. Uncertainties in the relative spectral radiance of the W lamp used at IC, and calibrated by the National Physical Laboratory (NPL), do not exceed $\pm$1.4 \% between $410$ nm and $800$ nm, and rise to $\pm$2.8 \% at 300 nm. 

\begin{table*}
\centering
\caption{FTS spectra taken for BF measurements.}
\begin{minipage}{180mm}
\begin{tabular}{lcccll}
\hline
Spectrum & Wavenumber\footnote{The equivalent wavelength ranges are: Spectrum A: $\lambda$ = 3920.5 \AA to 12191.8 \AA, Spectrum B: $\lambda$ = 2816.1 \AA to 4998.6 \AA.}        & Detector      & Filter     & Resolution  & Spectrum Filename\footnote{The named spectra were coadded to improve the signal-to-noise ratio of weak lines.} \\
         & Range (cm$^{-1}$) &               &            & (cm$^{-1}$) & \\
\hline
A (NIST) & 8200 - 25500        & Si photodiode & None     & 0.02   & Fe080311.001 to .003; Fe080411\_B.001 to .003 (220 coadds) \\
B (IC)   & 20000 - 35500       & Hammamatsu R11568 PMT    & Schott BG3 & 0.037 & Fe130610.002 to .047 (96 coadds)\\
\hline
\end{tabular}

Note: The identification of commercial products does not imply recommendation or endorsement by the National Institute of Standards and Technology, nor does it imply that the items identified are necessarily the best available for the purpose.
\end{minipage}
\label{table:spectra}
\end{table*}

Many of the upper energy levels studied here are linked to transitions that produced spectral lines contained entirely within the range of either spectrum A or B. In these cases, all branching fractions pertaining to those levels were derived from a single spectrum. Where lines associated with a given upper level spanned both spectra, their intensities were put on a common relative scale by comparing the intensity of lines in the overlap region between the two spectra. The process of intensity calibrating overlapping spectra is discussed in detail in \cite{ref:pickering01} and \cite{ref:pickering01b}.

For each target upper level, the predicted transitions to lower levels were obtained from the semi-empirical calculations of \cite{ref:kurucz07}. Emission lines from these transitions were then identified in our Fe spectra, and the \verb|XGremlin| package \citep{ref:xgremlin} used to fit Voigt profiles to those that were observed above the noise limit. The residuals from each fit were examined to ensure that the observed line profiles were free from self-absorption and not blended with other features. 

The spectra and fit results from \verb|XGremlin| were then loaded into the \verb|FAST| package \citep{ref:ruffoni13}, where the BFs for each observed target line were measured. Lines which were too weak to be observed -- typically those predicted by \cite{ref:kurucz07} to contribute less than 1 \% of the total upper level BF -- were not considered, as were lines that were either blended or outside the measured spectral range. Their predicted contribution to the total BF was assigned to a `residual' value, which was used to scale the sum over $l$ of the measured line intensity, $I_{ul}$, in Equation \ref{eqn:bf}. 

The calculation of experimental uncertainties in BFs measured by FT spectroscopy with \verb|FAST| has been discussed in our recent papers \citep{ref:ruffoni13b,ref:ruffoni13}. The uncertainty in a given BF, $\Delta\mbox{BF}_{ul}$, is

\begin{equation}
\Biggl(\frac{\Delta\mbox{BF}_{ul}}{\mbox{BF}_{ul}}\Biggr)^2 = (1 - 2\mbox{BF}_{ul})\Biggl(\frac{\Delta I_{ul}}{I_{ul}}\Biggr)^2 + \sum_{j=1}^n\mbox{BF}_{uj}^2\Biggl(\frac{\Delta I_{uj}}{I_{uj}}\Biggr)^2,
\end{equation}
where $I_{ul}$ is the calibrated relative intensity of the emission line associated with the electronic transition from level $u$ to level $l$, and $\Delta I_{ul}$ is the uncertainty in intensity of this line due to its measured signal-to-noise ratio and the uncertainty in the intensity of the standard lamp. From Equation \ref{eqn:trprob}, it then follows that the uncertainty in $A_{ul}$ is

\begin{equation}
\Biggl(\frac{\Delta A_{ul}}{A_{ul}}\Biggr)^2 = \Biggl(\frac{\Delta \mbox{BF}_{ul}}{\mbox{BF}_{ul}}\Biggr)^2 + \Biggl(\frac{\Delta \tau_{ul}}{\tau_{ul}}\Biggr)^2~,
\end{equation}
where $\Delta \tau_{ul}$ is the uncertainty in our measured upper level lifetime. Finally, the uncertainty in $\log(gf)$ of a given line is

\begin{equation}
\Delta \log(gf) = \log\Biggl(1 +  \frac{\Delta A_{ul}}{A_{ul}}\Biggr)~,
\end{equation}

\subsection{Upper Level Radiative Lifetimes}
\label{section:lif}
Radiative lifetimes are measured to $\pm 5$ \% using time-resolved laser-induced fluorescence (LIF) on an atomic beam of iron atoms. A diagram of the apparatus is shown in \cite{ref:obrian91}. The beam is produced by sputtering iron atoms in a hollow cathode discharge.  The electrical discharge is operated in ${\approx}50$ Pa argon gas.  A DC current of ${\approx}30$ mA maintains the discharge between ${\approx}10$ A, $10$ $\umu$s duration pulses at $30$ Hz repetition rate. The hollow cathode, which is lined with a foil of pure iron, is closed on one end except for a $1$ mm hole which is flared on one side to act as a nozzle. Energetic argon ions accelerated through the cathode fall potential efficiently sputter the iron from the surface of the cathode.  The iron atoms (neutral as well as singly-ionized) are differentially pumped through the nozzle amidst a flow of argon gas into a low pressure ($10^{-2}$ Pa) scattering chamber.  This "beam" is slow (neutrals are moving ${\sim}5\times 10^4$ cm/s and ions somewhat faster) and weakly collimated. 

Measurement of the odd-parity level lifetime required single-step laser excitation.  In this technique the atomic beam is intersected at right angles by a single beam from a nitrogen laser-pumped dye laser $1$ cm below the nozzle. The delay between the discharge pulse and the laser pulse is adjustable, and optimized typically at ${\sim}20$ $\umu$s, which corresponds to the average transit time of the iron atoms. The scattering volume is at the center of a set of Helmholtz coils which zeroes the magnetic field to within $\pm 2$ $\umu$T.  This very low field ensures that the excited iron atoms do not precess about the Earth's magnetic field, thus eliminating the potential for Zeeman quantum beats in the fluorescence. The dye laser is tunable over the range $205$ nm to $720$ nm using a large selection of dyes as well as frequency doubling crystals.  It has a bandwidth of ${\sim}0.2$ cm$^{-1}$, a half-width duration of ${\sim}3$ ns and, more importantly for this work, terminates completely in a few ns. The laser allows for selective excitation of the level under study, eliminating the problem of cascade from higher-lying levels that plagued earlier, non-selective techniques. The laser is tuned to a transition between the ground state or a low-lying metastable level and the level under study.  Identifying the correct transition is non-trivial, particularly for a dense, line rich spectrum such as Fe~I and Fe~II.  The laser is tuned to within ${\approx}0.1$ nm of the transition by adjusting the angle of the grating, which is the tuning element of the laser, while measuring the wavelength with a $0.5$ m monochromator.  A LIF spectrum of $0.5$ nm to $1.0$ nm range is then recorded using a boxcar averager by slowly changing the pressure in an enclosed volume surrounding the grating.  Pressure scanning provides exceptional linearity and reproducibility.  The pressure scanned spectrum is then compared to the published linelist from the NIST database\footnote{http:// physics.nist.gov/PhysRefData/ASD/index.html} to correctly identify the line of interest.

Fluorescence is collected in a direction mutually orthogonal to the atomic and laser beams through a pair of fused-silica lenses comprising an f/1 optical system.  A spectral filter, either a broadband colored-glass filter or a narrowband multilayer dielectric filter, is inserted between the two lenses where the fluorescence is approximately collimated.  The filter is chosen to maximize fluorescence throughput while reducing or eliminating scattered laser light and eliminating possible cascade from lower-lying levels. Fluorescence is focused onto the photocathode of a RCA 1P28A photomultiplier tube (PMT) and the PMT signal is recorded using a Tektronix SCD1000 transient digitizer. The bandwidth of the PMT, digitizer and associated electronics is adequate to measure lifetimes down to ${\sim}2$ ns.  The lifetimes reported here are in the $10$ ns to $25$ ns range and are well within the bandwidth limits. The characteristics of this PMT, i.e. fast rise time and high spectral response in the UV and visible, are favourable for radiative lifetime measurements.

The digitizer is triggered with the signal from a fast photodiode which is illuminated by light picked off from the nitrogen laser. Recording of the fluorescence by the digitizer is delayed until after the dye laser pulse has completely terminated, making deconvolution of the laser temporal profile and fluorescence signals unnecessary.  Each data record consists of an average of $640$ fluorescence decays followed by an average of $640$ background traces with the laser tuned off-line.  The data is divided into an early time and a late time interval for analysis.  A linear least-square fit to a single exponential is performed on the background subtracted fluorescence decay to determine a lifetime for each interval.  Comparison of the lifetimes in the two intervals is a sensitive indicator of whether the decay is a clean exponential or whether some systematic effect has rendered it non-exponential.  Five of these decay times are averaged together to determine the lifetime.  The lifetime of each odd-parity level is measured twice, using two different laser transitions.  This redundancy helps to ensure that the transition is classified correctly, free from blends, and is identified correctly in the experiment.

Measurement of the even-parity levels reported here required two-step laser excitation.  The introduction of a second laser results in an added layer of complexity in the excitation of the level and timing of the experiment, as well as more stringent requirements for the filtering of the fluorescence. The fluorescence detection, recording and analysis is identical to the one-laser experiment.  While it is possible to pump two dye lasers using one nitrogen laser, this limits the power available in either laser beam. Instead, we used two dye lasers each with its own nitrogen laser pump.  The delay generator which, in the one laser experiment is used to trigger the laser ${\sim}20$ $\umu$s after the discharge pulse, is in this case used to trigger a second dual gate generator that has very precise timing ($\pm 1$ ns) between its two gates.  These gates are used to trigger the two nitrogen lasers which pump the dye lasers. Because the two nitrogen lasers have different thyratron charging and firing mechanisms, there is a substantial amount of timing jitter (approximately $\pm 20$ ns) between the resulting dye laser pulses.  This jitter results in some additional shot-to-shot fluctuation in the final measurement as the population in the intermediate level has decayed more or less from its peak.  The lifetimes of all the intermediate levels used but one is substantially longer than this jitter ($60$ ns to $85$ ns as measured by \cite{ref:obrian91}), so the added shot-to-shot noise was not severe. Even the measurement with the short-lived ($9.6$ ns as measured by \cite{ref:obrian91}) intermediate level had only ${\sim}2$ \% statistical scatter in the final average.  The delay between the two lasers is adjusted such that the laser which drives the transition from the intermediate level to the even-parity level being studied (laser 2) arrives on average ${\sim}20$ ns after that which drives a transition between the ground or low-lying metastable level and an intermediate odd-parity level (laser 1). The trigger signal for the boxcar and digitizer was from the fast photodiode illuminated with light from the laser 2 nitrogen laser.  

\begin{table*}
  \centering
    \caption{Radiative lifetimes for Fe~I levels used to derive the $\log(gf)$s in Table \ref{table:results}. The uncertainty in both our lifetime measurements and those of O'Brian et al. (1991) was $\pm 5$ \%.}
    \begin{minipage}{180mm}
      \begin{tabular}{lcccccccccc}
\hline
Configuration & Term & J & \multicolumn{2}{c}{Upper Level}       & Intermediate & \multicolumn{2}{c}{Laser Wavelengths\footnote{Laser wavelengths are from the NIST Atomic Spectra Database (http://www.nist.gov/pml/data/asd.cfm).}} & Observation & Our & Previous \\
\cline{4-5}
\cline{7-8}
              &      &   &             &      & Level        & Step 1 & Step 2                                                                                                                                    & Wavelength\footnote{Fluorescence was observed through $\sim 10$ nm bandpass multi-layer dielectric filters.  The filter angle was adjusted where needed to centre the bandpass at the indicated wavelength.}  & Lifetime & Lifetime\\
              &      &   & (cm$^{-1}$) & (eV) & (cm$^{-1}$)  & \multicolumn{2}{c}{(nm)}                                                                                                                          & (nm)        & (ns) $\pm 5$ \% & (ns) $\pm 5$ \%\\
\hline
\multicolumn{10}{l}{Radiative lifetime for odd parity Fe~I levels using single step excitation.} \\
\hline
$3d^6$($^3$H)$4s4p$($^3$P$^o$) & y$^1$G$^o$ & 4 & 48702.532 & 6.0383441 &            & 241.906,&         &     & 21.0 \\
                               &            &   &           &           &            & 414.341 &         &     &      \\
\hline                               
\multicolumn{10}{l}{Radiative lifetimes for even parity Fe~I levels using 2-step excitation.} \\
\hline
$3d^6$($^5$D)$4s$($^6$D)$5s$   & e$^5$D     & 4 & 44677.003 & 5.5392422 & 25899.987  & 385.991 & 532.418 & 561 & 15.4 & 15.6 $\pm$ 0.9\footnote{\cite{ref:marek79}.}\\
$3d^7$($^4$F)$5s$              & e$^5$F     & 4 & 47377.952 & 5.8741171 & 26140.177  & 382.444 & 470.727 & 490 & 19.9 \\
$3d^7$($^4$F)$5s$              & e$^3$F     & 4 & 47960.937 & 5.9463981 & 26140.177  & 382.444 & 458.151 & 600 & 22.1 \\
$3d^6$($^5$D)$4s$($^4$D)$5s$   & e$^3$D     & 3 & 51294.217 & 6.3596721 & 26140.177  & 382.444 & 397.438 & 500 & 9.9 \\
$3d^6$($^5$D)$4s$($^4$D)$5s$   & g$^5$D     & 3 & 51770.554 & 6.4187304 & 26140.177  & 382.444 & 390.052 & 408 & 12.1 \\
$3d^6$($^5$D)$4s$($^4$D)$5s$   & e$^3$D     & 1 & 52039.889 & 6.4521236 & 26339.694  & 385.637 & 388.992 & 500 & 10.4 \\
$3d^6$($^5$D)$4s$($^6$D)$4d$   & e$^5$P     & 2 & 52067.466 & 6.4557907 & 26140.177  & 382.444 & 385.585 & 446 & 14.3 \\
$3d^7$($^4$F)$4d$              & f$^3$F     & 4 & 54683.318 & 6.7798670 & 36686.174  & 404.581 & 555.489 & 570 & 23.5 \\
\hline                               
\multicolumn{10}{l}{Additional radiative lifetimes from O'Brian et al. (1991).} \\
\hline
$3d^64s$($^6$D)$5s$            & e$^7$D     & 4 & 43163.323 & 5.3515698 &            & 423.59 & & & & 8.5  \\
                               &            &   &           &  &            & 427.12 &      \\
$3d^64s$($^6$D)$5s$            & e$^7$D     & 2 & 43633.530 & 5.4098680 &            & 418.70 & & & & 8.4  \\
                               &            &   &           &  &            & 423.36 &      \\
$3d^64s$($^6$D)$4d$            & f$^5$F     & 4 & 51461.667 & 6.3804332 &            & 315.32 & & & & 12.7 \\
\hline
\end{tabular}

Note: The configuration, term, and energy level data are taken from \cite{ref:nave94}.
\end{minipage}
\label{table:lifetimes}
\end{table*}

The two lasers are sent through the scattering chamber at slight angles relative to each other, such that they intersect in the viewing volume.  Once laser 1 is tuned onto the appropriate transition to drive the intermediate level, it is left there for the duration of the measurement. A narrowband, multilayer dielectric filter is inserted in the collection optics which completely blocks fluorescence from the intermediate level but transmits fluorescence from the upper level.  Laser 2 was tuned on and off the transition to provide the fluorescence and background traces as in the one-step experiment. The fluorescence was observed to go away when either laser 1 or laser 2 was blocked and the other laser was allowed to pass through the system, ensuring that  it was indeed from a two-step process.  This provides the assurance that the correct lifetime is being measured that a redundant measurement gives in the one-step experiment.  Each two-step lifetime was therefore measured only once.

Systematic effects such as Zeeman quantum beats and bandwidth limits are well-studied and controlled in the experiment.  Another effect, the flight out of view effect, is caused by atoms leaving the viewing volume before fluorescing.  This effect is only a problem for long lifetimes, greater than $300$ ns for neutrals and greater than $100$ ns for ions, and is not a problem for the current set of lifetimes.  In addition to understanding and minimizing these systematics, we also regularly measure a set of benchmark lifetimes, to compare our measured values to the known lifetimes.  These benchmarks are lifetimes that are either very well known from theoretical calculations, or from an experiment which has significantly smaller and generally different systematic uncertainties from our own.  For the current set of lifetimes, we measured three benchmarks which approximately bracketed the range of values reported here. These are: $2p~^2P_{3/2}$  level of singly ionized Be at 8.8519(8) ns (variational method calculation  \citep{ref:yan98}); the $3p~^2P_{3/2}$ level of neutral Na at 16.23(1) ns (accuracy of $\le 0.1$ \% at 90 \% confidence level) taken from the recent NIST critical compilation of \cite{ref:kelleher08}; and the $2p_2~4p'[1/2]_1$ level of neutral Ar at 27.85(7) ns (beam-gas-laser-spectroscopy \citep{ref:volz98}).  Benchmarks are measured in exactly the same way as the Fe~I lifetimes except that the cathode lining is changed in the cases of the Be$^{+}$  and Na measurements. With these benchmarks we are able to quantify and make small corrections for any residual systematic effects ensuring that our final results are well within the stated uncertainty of $\pm 5$ \%. A recent comparison of LIF measurements in Sm II by \cite{ref:lawler08} suggests that the $\pm 5$ \% is a conservative estimate of the lifetime uncertainty.

The lifetime results are given in Table \ref{table:lifetimes}. A total of 1 odd-parity and 8 even-parity level lifetimes were measured; most for the first time. The even-parity $e^5D_4$ level at $44677.003$ cm$^{-1}$ was also measured by \cite{ref:marek79} using delayed coincidence detection after laser excitation, and agrees with our lifetime to about $\pm 1$ \%. This good level of agreement is what we have come to expect between modern, laser-based methods. 

\section{Results}\label{section:results}
Table \ref{table:lifetimes} lists the Fe~I upper levels that were targeted in this study. They were selected because their branches to lower levels produce many spectral lines of interest to the GES survey that currently have either no experimentally measured $\log(gf)$ value in the literature, or a $\log(gf)$ known to worse than $\pm 25 \%$. We also included two levels, those at $43633.530$ cm$^{-1}$ and $51461.667$ cm$^{-1}$, for which accurate lifetimes and $\log(gf)$s were reported by \cite{ref:obrian91}. These served primarily as a means to check the accuracy of $\log(gf)$s produced with the aid of the \verb|FAST| code, but in remeasuring them we were also able to improve upon the experimental uncertainty achieved by \cite{ref:obrian91} and provide $\log(gf)$s for a number of weaker branches not included in their paper. Some further lines reported by \cite{ref:obrian91} appear in branches from other upper levels, as do a few lines for which accurate $\log(gf)$s were reported by \cite{ref:blackwell82} and \cite{ref:bard91}. Again, these served as a means to check the accuracy of our results.

Our measured branching fractions, transition probabilities, and $\log(gf)$s are listed in Table \ref{table:results} along with the most accurate $\log(gf)$s previously available in the literature. The lower level terms, and transition vacuum wavenumbers and air wavelengths were taken from \cite{ref:nave94}, where possible. For the small number of lines not included in \cite{ref:nave94}, the transition vacuum wavenumber and air wavelength shown were obtained from our FT spectra by calibrating the measured wavenumber scale to match the calibrated scale used by \cite{ref:nave94}. These lines are marked in Table \ref{table:results} by a `*' in the Lower Level column. 

Table \ref{table:results} is sorted in order of ascending transition wavenumber, with lines grouped by common upper level energy. For each set of lines, the upper level energy, configuration, term, and J value, and measured lifetime are given as a header row. The unobserved `residual' BF, described in Section \ref{section:bf}, is given in the BF column at the end of each set. The lines that contribute to these residuals are given in Table \ref{table:missing} where they have either been observed in previous studies, or predicted by \cite{ref:kurucz07} to contribute more than 1 \% to the total BF. Reasons for their omission in this study are given.

For six of the eleven upper levels (those at 43633.530 cm $^{-1}$, 44677.003 cm $^{-1}$, 47377.952 cm $^{-1}$, 47960.937 cm $^{-1}$, 48702.532 cm $^{-1}$, and 51294.217 cm $^{-1}$) the residual BF amounted to less than $5 \%$, and arose solely from lines predicted by \cite{ref:kurucz07} to contribute to the total set of branches that were too weak to be observed experimentally. For the remaining five levels (those at 51461.667 cm $^{-1}$, 51770.554 cm $^{-1}$, 52039.889 cm $^{-1}$, 52067.446 cm $^{-1}$, and 54683.318 cm $^{-1}$) a large majority of branches were observed, but at least one stronger line was unavailable due to being unobserved above the spectral noise, blended with another line, or significantly separated in wavenumber from the rest of the branches (which prevents correct intensity calibration). In all cases, the missing BF was taken from previously published values, if they existed, or from Kurucz's calculations otherwise, as shown in Table \ref{table:missing}. Any error in these values will affect the overall normalisation of $\log(gf)$s for the level in question, in turn leading to a systematic error in their value. However, we expect this error to be small, and so have neglected it, for two reasons. Firstly, there is good agreement between our $\log(gf)$s and those from \cite{ref:obrian91} for branches from the $51461.667$ cm$^{-1}$ level, which has a residual BF of 0.124 (the largest of all levels) and secondly, this residual can be varied by as much as $\pm 20 \%$ without the normalisation error exceeding the random uncertainty in $\log(gf)$ of any of the branches. 

Lines that are of particular interest for the GES survey are marked in Table \ref{table:results} in the ``GES Target?" column. In some cases the $\log(gf)$s for these lines have been measured in earlier studies, in which case we have sought to reduce their uncertainty. For lines originally measured by \cite{ref:may74}, the quoted published $\log(gf)$s are the corrected values given by \cite{ref:fuhr06} in their recent critical compilation of Fe~I $\log(gf)$s. In preparing their compilation, these authors noted that the lifetimes used by \cite{ref:may74} originate from data produced in the 1960s and early 1970s. Comparing these to the cascade-free LIF lifetimes measured by \cite{ref:obrian91}, they found that for 13 energy levels between $52000$ cm$^{-1}$ and $57000$ cm$^{-1}$ the lifetimes given by \cite{ref:obrian91} were systematically shorter by about $20 \%$, most likely due to the absence of cascade effects. For levels below $36000$ cm$^{-1}$, this systematic error vanished. \cite{ref:fuhr06} therefore corrected the $\log(gf)$s given by \cite{ref:may74} for levels above $36000$ cm$^{-1}$ to make them consistent with the lifetime data of \cite{ref:obrian91}. For the remaining $\log(gf)$s, \cite{ref:fuhr06} found fair agreement with the results of \cite{ref:obrian91} and \cite{ref:blackwell82} where they overlapped. However, the scatter was ``quite large", suggesting that the uncertainties given by \cite{ref:may74} should be significantly larger. In Table \ref{table:results}, the uncertainties in $\log(gf)$s from \cite{ref:may74} are therefore given as a letter `D' or `E' to follow the notation used by \cite{ref:fuhr06}. A letter `D' indicates that the uncertainty is likely to be up to $50 \%$, whereas an `E' indicates a probable uncertainty greater than $50 \%$, but within a factor of two in most cases.

\begin{figure}
\centering
\includegraphics[scale=0.48]{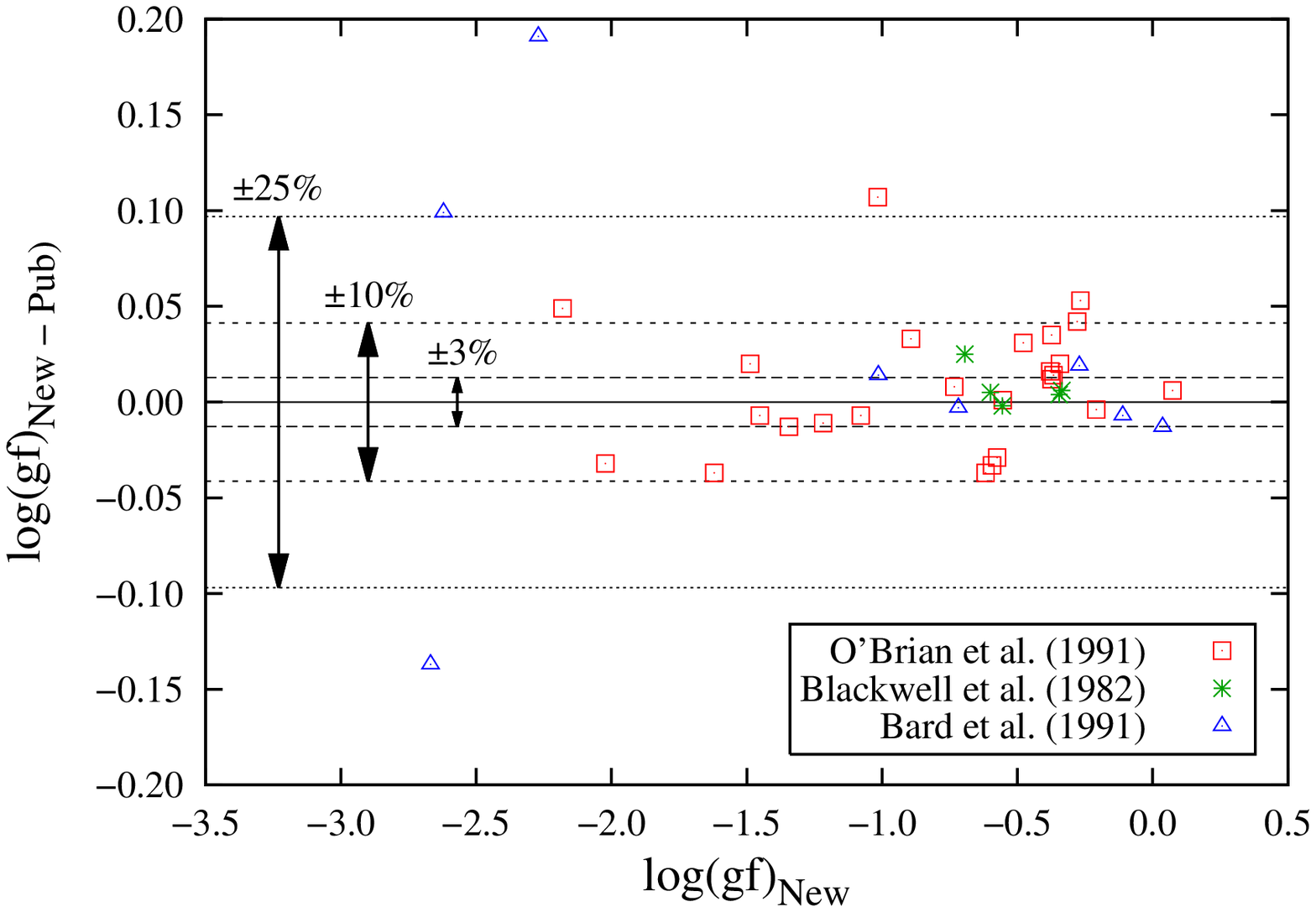}
\includegraphics[scale=0.48]{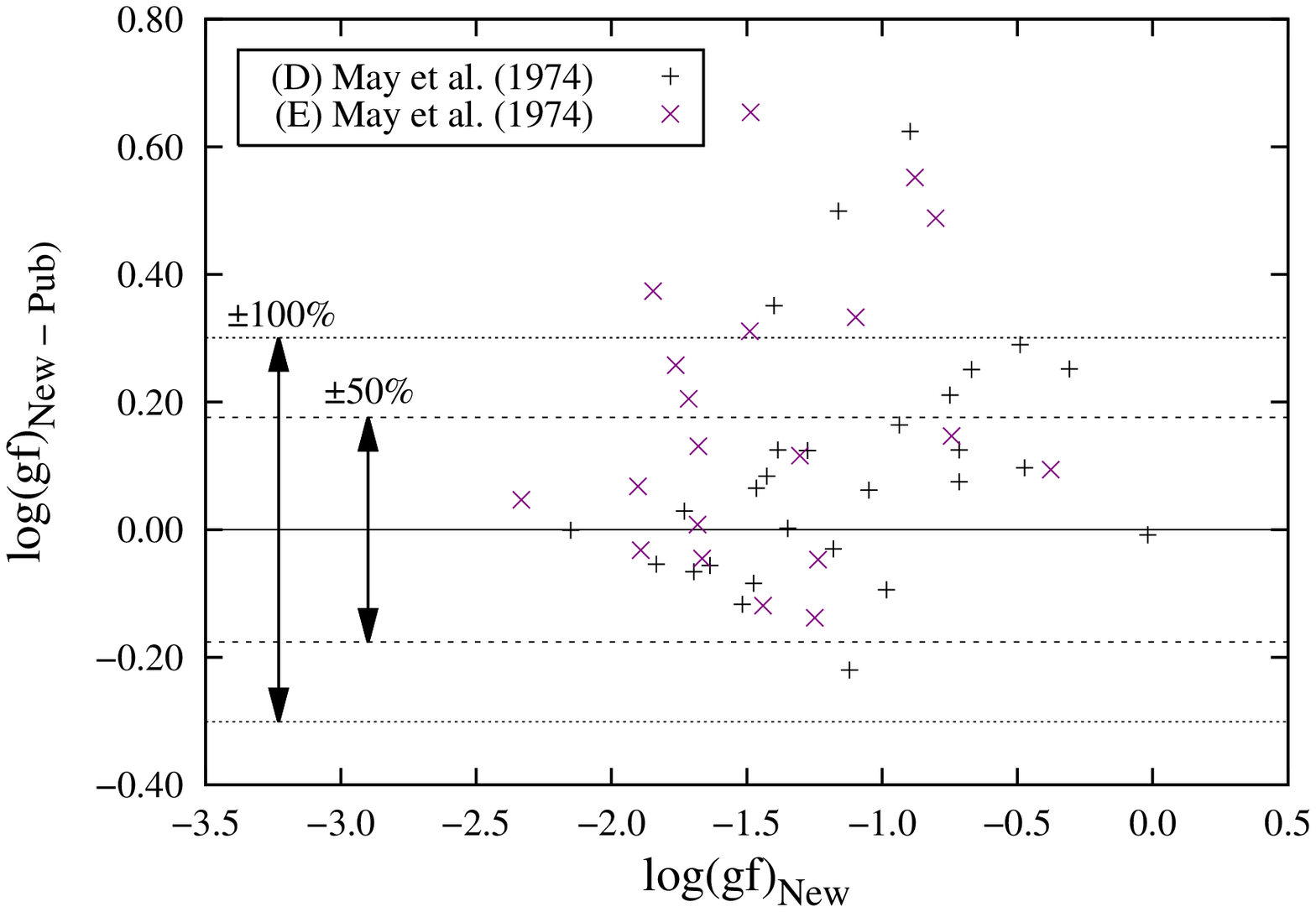}
\caption{A comparison between the log(gf)s of this work and those already in the literature. The upper panel shows results from O'Brian et al. (1991), Blackwell et al. (1982), and Bard et al. (1991), which agree well with our new log(gf) values. The lower pane shows the corrected results from May et al. (1974), which have a considerably lower accuracy.}
\label{fig:comparepub}
\end{figure}

Figure \ref{fig:comparepub} shows a comparison between our new $\log(gf)$s and those published previously. The top panel shows the difference between our values and those reported by \cite{ref:obrian91}, \cite{ref:blackwell82}, and \cite{ref:bard91}. The long dashed, short dashed and dotted horizonatal lines indicate uncertainties of $\pm 2$ \%, $\pm 10$ \% and $\pm 25$ \%, respectively, corresponding to uncertainties coded ` A', `B' and `C' by \cite{ref:fuhr06}. The work of \cite{ref:blackwell82} continues to serve as a gold-standard for Fe~I $\log(gf)$s in the literature. Five lines from their study are also included our work, and the $\log(gf)$ for each agrees within their combined experimental uncertainty of $\pm 5 \%$. There is also very good agreement with the results of \cite{ref:obrian91} and \cite{ref:bard91}. 25 of the 29 $\log(gf)$s from these papers agree within the combined experimental uncertainties with no discernible systematic offset between the published results and our new values. Together, these testify to the general accuracy of our $\log(gf)$ measurements and the accuracy of the \verb|FAST| code in extracting $\log(gf)$s from FT spectra. 

The lower panel of Figure \ref{fig:comparepub} shows the difference between our values and the corrected $\log(gf)$s given by \cite{ref:fuhr06} for the data reported by \cite{ref:may74}. The dashed and dotted lines this time indicate uncertainties of $\pm 50$ \% and $\pm 100$ \%, respectively, which correspond to uncertainties coded `D' and `E' by \cite{ref:fuhr06}. 31 of the 39 corrected $\log(gf)$s from \cite{ref:may74} agree with our new values when considering the enlarged uncertainties attributed to them by \cite{ref:fuhr06}, but there is considerable scatter in the results, as was also noted by \cite{ref:fuhr06}. There is also a systematic offset of $\log(gf)_{(New - Pub)} = 0.12$ for these lines. Our new $\log(gf)$s for these lines are accompanied by considerably smaller uncertainties; typically less than $25 \%$, with some as low as $5 \%$ for stronger lines.

\section{Impact on Solar Spectral Synthesis}\label{section:impact}
The Sun offers an excellent test-bed for new atomic data, with its high-resolution spectrum \citep{ref:kittpeak} and accurately known fundamental parameters \citep{ref:almanac}. To assess the impact of our new $\log(gf)$s on stellar syntheses, and also verify their general accuracy, we have determined line-by-line solar Fe abundances for a subset of 36 lines listed in Table \ref{table:results} using both our new $\log(gf)$s and the best previously published values that are not of astrophysical nature. These lines, shown in Table \ref{table:solar}, were selected as they are blend-free at the spectral resolution of the Kitt Peak Fourier Transform Spectrometer ($R \approx 200000$) flux atlas \citep{ref:kittpeak}, and are accompanied by good broadening parameters and accurate continuum placement. The synthesis and abundance determination were performed under the assumption of local thermodynamic equilibrium (LTE), with the one dimensional, plane-parallel radiative transfer code SME \citep{ref:valenti96}, using a MARCS model atmosphere \citep{ref:gustafsson08}. 


We adopted a solar effective temperature $T_{\rm eff}=5777\,K$, a surface gravity $\log(grav.)=4.44$, a microturbulence of $\xi_{\rm vmic}=1.0\rm\,km\,s^{-1}$, and a projected rotational velocity of $v_{\rm rot}\sin(i)=2.0\rm\,km\,s^{-1}$. The radial-tangential macro-turbulence velocity, $\xi_{\rm vmac}$, was varied between $1.5\rm\,km\,s^{-1}$ and $2.5\rm\,km\,s^{-1}$ to match the observed profile. The instrumental profile was assumed to be Gaussian. The line profiles were fitted individually using $\chi^2$-minimisation between observed and synthetic spectra and varying the iron abundance. 

\begin{figure}
\centering
\includegraphics[scale=0.48]{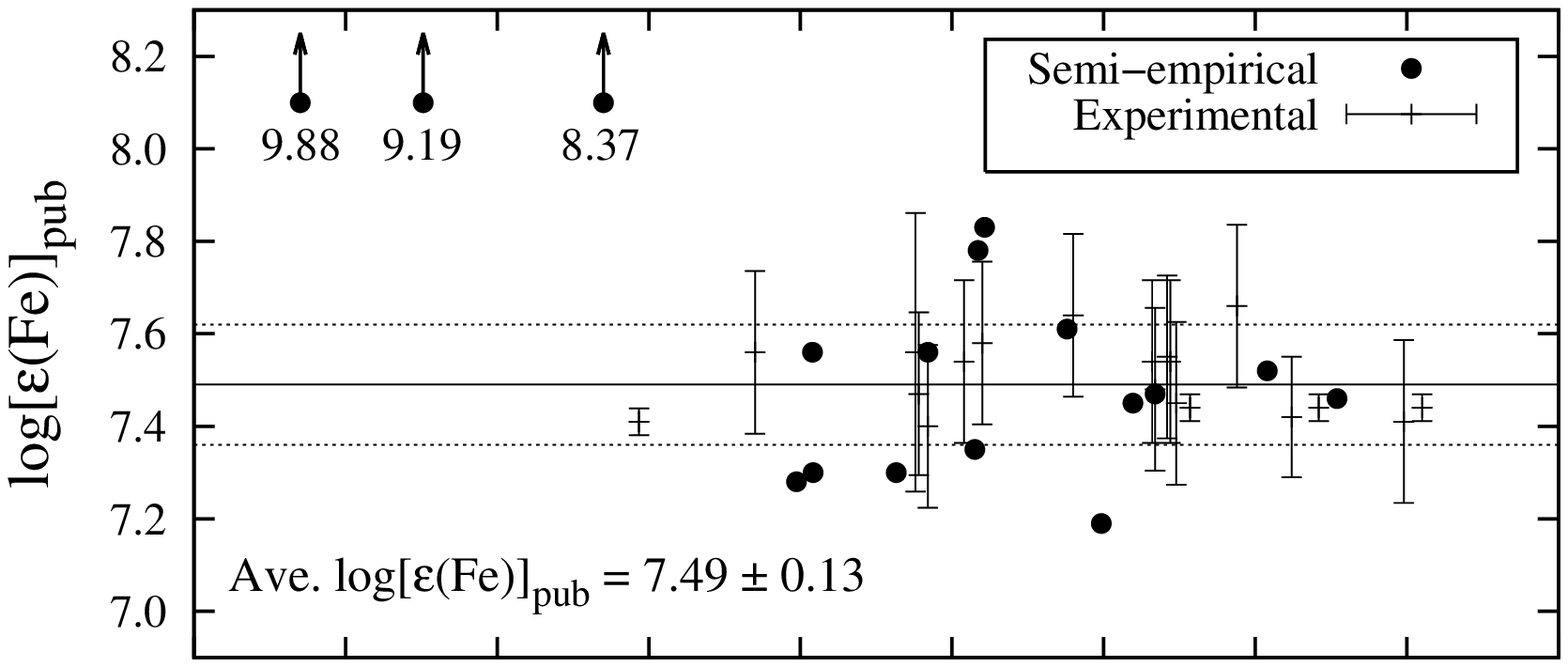} \\
\includegraphics[scale=0.48]{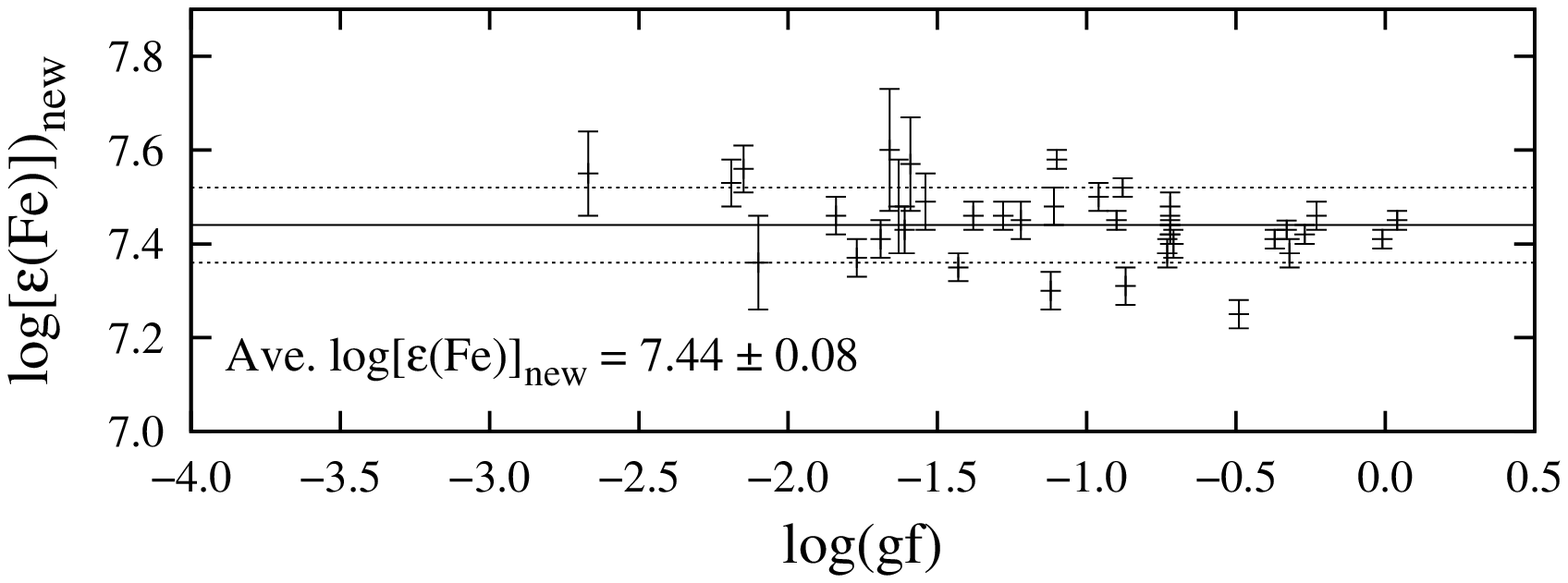} \\
\caption{Solar metalicity, $\log[\epsilon(\mbox{Fe})]$, obtained from the synthesis of individual lines listed in Table \ref{table:results} using the $\log(gf)$ values from this work, $\log(gf)_{\mbox{New}}$, and the best previously published values, $\log(gf)_{\mbox{Pub}}$. Only those lines that appear unblended in the solar spectrum are included. The three points at the top-left of the upper pane are discrepant semi-empirical values that lie outside the plotted range at the values shown. They were not included in the calculation of the average $\log[\epsilon(\mbox{Fe})]_{\mbox{Pub}}$.}
\label{fig:solar}
\end{figure}

The results are shown in Figure \ref{fig:solar}, where the abundances are plotted as a function of $\log(gf)$ on the standard astronomical scale.

\begin{equation} 
\log[\epsilon(\mbox{Fe})] = \log_{10}\Bigl[\frac{N(\mbox{Fe})}{N(\mbox{H})}\Bigr] + 12~,
\label{eqn:abundance}
\end{equation} 
where $N(\mbox{Fe})$ and $N(\mbox{H})$ are the number of iron and hydrogen atoms per unit volume, respectively. 

\begin{figure*}
\centering
\includegraphics[scale=0.59]{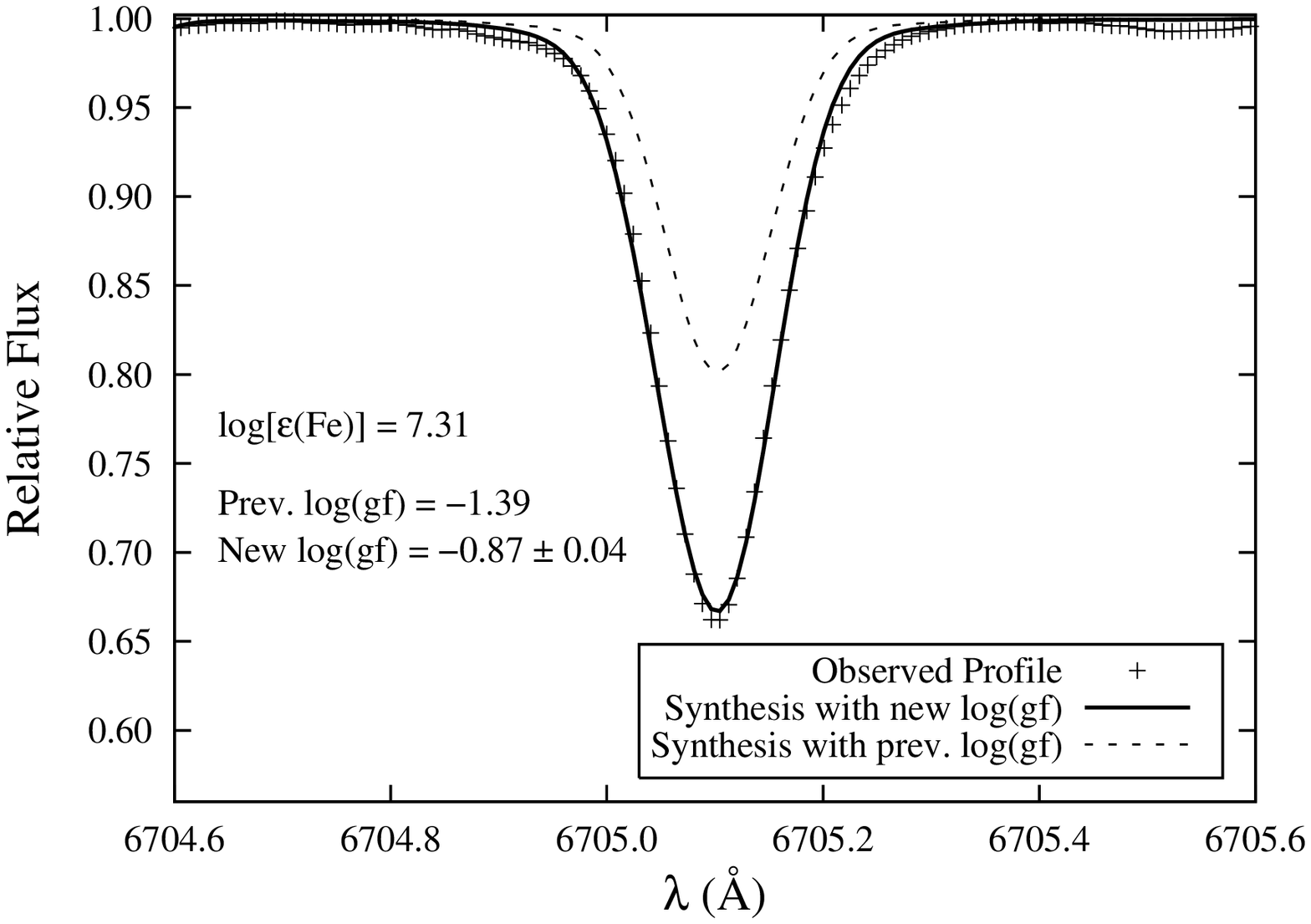} \\
\includegraphics[scale=0.59]{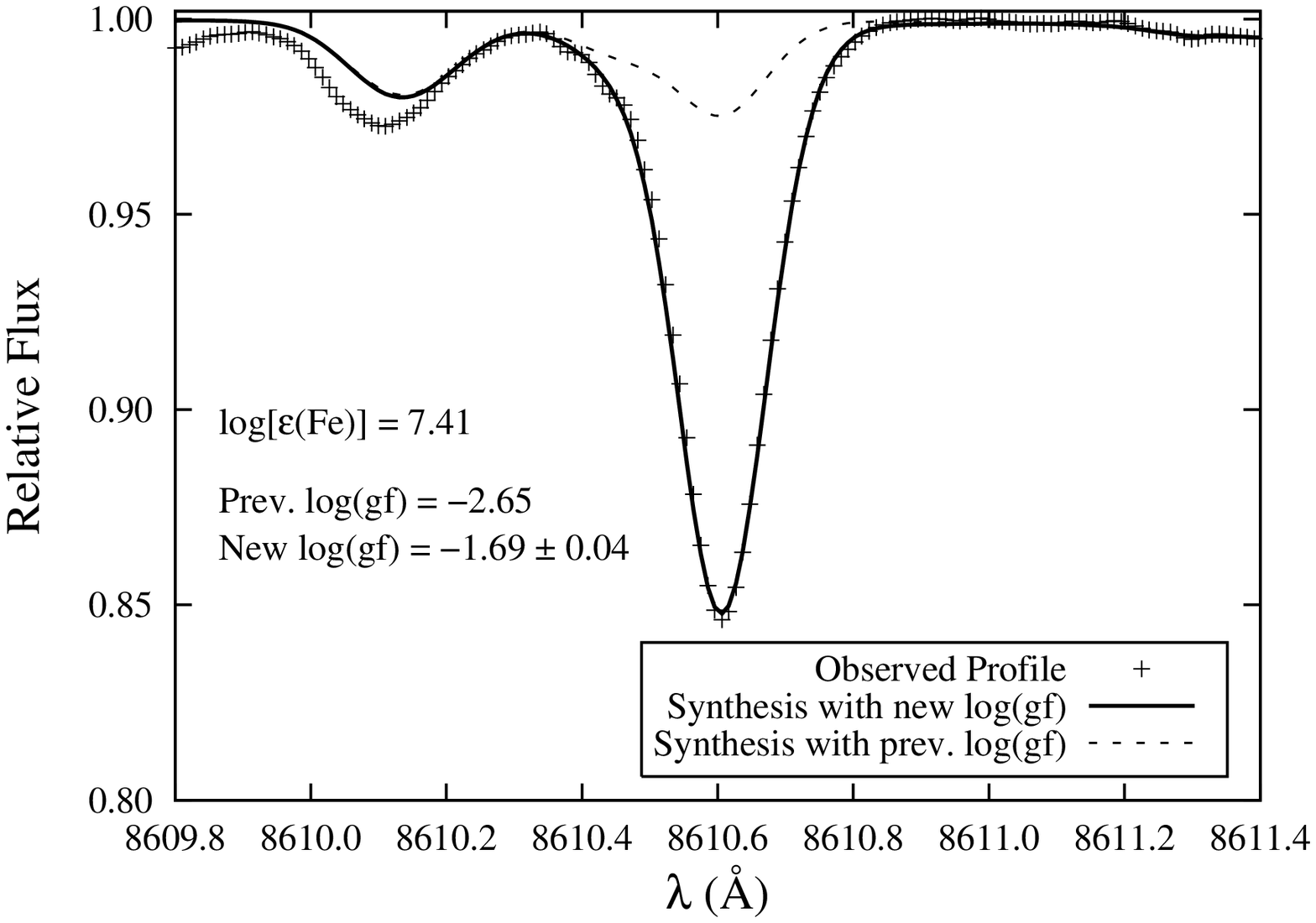} \\
\includegraphics[scale=0.59]{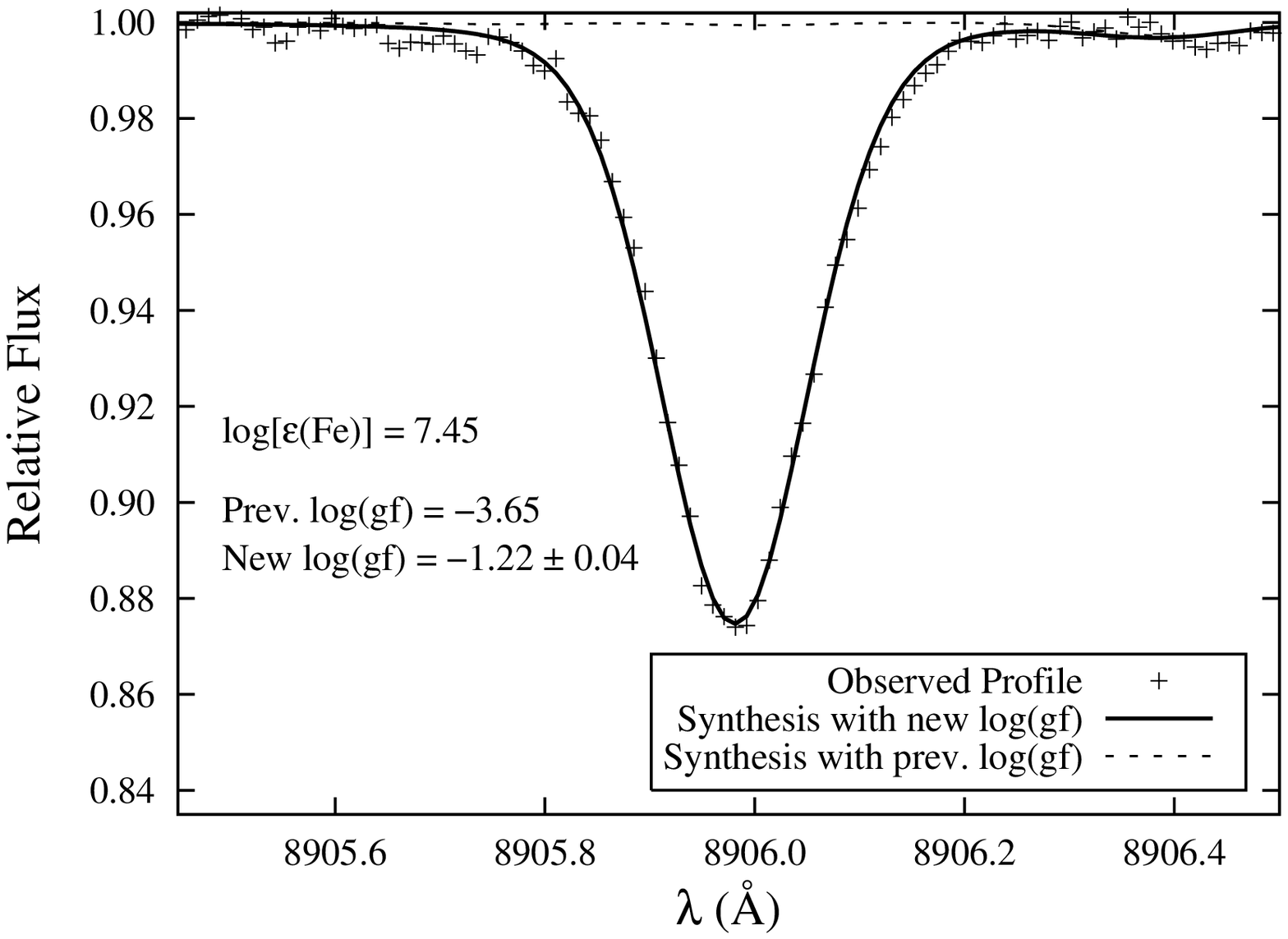}
\caption{Three examples of lines observed in the solar spectrum and synthesised using both the $\log(gf)$s from this work and the best previously published values. The solid lines show the synthetic profiles obtained using the $\log(gf)$s from this work and the quoted values of $\log[\epsilon(\mbox{Fe})]$. The dashed lines show the synthetic profiles obtained using the same values of $\log[\epsilon(\mbox{Fe})]$, but adopting the best previously published $\log(gf)$s, shown in Table \ref{table:solar}.}
\label{fig:lines}
\end{figure*}

Reassuringly, the new experimental data result in a small line-to-line scatter (0.08 dex)\footnote{The unit dex stands for decimal exponent. $x$ dex = $10^x$.} and a mean abundance of $7.44$, which is in good agreement with recent publications, such as $7.43 \pm 0.02$ from \cite{ref:bergemann12} (MARCS, LTE result). In contrast, the best previously published values (omitting the discrepant semi-empirical values shown in Figure 3) produce an abundance of $7.49 \pm 0.13$, with the significantly larger scatter driven by the lines with no previous laboratory measurements. The observed and best-fit synthetic profiles of three of these lines are shown in Figure \ref{fig:lines}. They all fall within the GES wavelength windows, and two of them are also in the near-infrared \emph{Gaia} Radial Velocity Spectrometer window \citep{ref:katz04}.  

Other lines with significant improvements in the solar modelling, but not shown in Table \ref{table:solar}, are those at 4079.2 {\AA}, 4933.9 {\AA}, and 5171.7 {\AA}, which are partly blended with astrophysically interesting lines such as the Ba~II 4934.0 {\AA} line, the Mn~I 4079.2 {\AA} line, and the Mg-I triplet line at 5172.7 {\AA}.  

\section{Summary}\label{section:summary}
In Table \ref{table:results}, we have provided new $\log(gf)$ values for 142 Fe~I lines from 12 upper levels, which include 38 lines of particular interest for the analysis of stellar spectra obtained by the GES survey. Where $\log(gf)$s existed for these lines in the literature, we have found good agreement with our new values, which in many cases have smaller experimental uncertainties than those previously reported. This is especially true for uncertainties in $\log(gf)$s from \cite{ref:may74}, which have been reduced from $50 \%$ or more to less than $25 \%$ in most cases.

This work represents part of an on-going collaboration between Imperial College London, U. Wisconsin, and NIST to provide the astronomy community with Fe~I $\log(gf)$ values needed for the analysis of astrophysical spectra. Further publications will follow in the near future.

\section{Acknowledgements}\label{section:acknowledge}
MPR and JCP would like to thank the UK Science and Technology Facilities Council (STFC) for supporting this research and the European Science Foundation (ESF), under GREAT/ESF grant number 5435, for funding international travel to discuss research plans with the wider GES team. EDH and JEL acknowledge the support of the US National Science Foundation (NSF) for funding the LIF lifetime measurements under grants AST-0907732 and AST-121105. KL acknowledges support by the European Union FP7 programme through European Research Council (ERC) grant number 320360.

Please note that the identification of commercial products in this paper does not imply recommendation or endorsement by the National Institute of Standards and Technology, nor does it imply that the items identified are necessarily the best available for the purpose.
                                                                                                                                                                                                                                                                                                            
\end{minipage}                                                                                                                                                                                                                                                                                                           
\label{table:solar}                                                                                                                                                                                                                                                                                                      
\end{table*}                                                                                                                                                                                                                                                                                                             

\twocolumn

\end{document}